\documentclass[aps,prl,twocolumn,superscriptaddress,longbibliography,footinbib]{revtex4-2}
\usepackage{mathrsfs}
\usepackage{epsfig}
\usepackage{graphicx}
\usepackage{amsfonts}
\usepackage[figuresright]{rotating}
\usepackage{amssymb}
\usepackage{amsmath}
\usepackage{dcolumn}
\usepackage{bm}
\usepackage{color}
\usepackage{xcolor}

\usepackage[pdftex,colorlinks=true]{hyperref}

\def\be{\begin{equation}} \def\ee{\end{equation}}
\def\bea{\begin{eqnarray}} \def\eea{\end{eqnarray}}

\newcommand{\pku}{State Key Laboratory of Artificial Microstructure and Mesoscopic Physics, School of Physics, Peking University, 100871 Beijing, China}

\newcommand{\affC}{Cavendish Laboratory, University of Cambridge, Cambridge CB3 0HE, United Kingdom}
\begin{document}
\title{Anomalous Floquet Heating from Sparse Long-Range Interactions}
\author{Chenyue Guo}
\affiliation{\pku}

\author{Andrea Pizzi}
 \affiliation{\affC}
 
\author{Hongzheng Zhao}
\email{hzhao@pku.edu.cn}
\affiliation{\pku}

\begin{abstract}
Regular lattices of interacting particles under a periodic drive typically heat with rate $\gamma \sim e^{-\mathcal{O}(\omega)}$ which is exponentially suppressed in drive frequency $\omega$. Here, we show that sparse infinite-range interactions, which have recently become accessible in quantum simulators, can lead to anomalous heating with rate $\gamma \sim e^{-\mathcal{O}(\sqrt{\omega})}$. This anomaly originates from the broad distribution of coordination numbers across the network: as the driving frequency increases, heating becomes dominated by sites with larger coordination numbers, making the characteristic local energy scale relevant for heating grow with frequency. For small-world networks, we develop an analytic theory that thoroughly matches our large scale numerics. Finally, we discuss how network topology can serve as a control knob for engineering non-equilibrium phases of matter. Our results uncover a new mechanism for Floquet heating and suggest new routes toward stabilizing nonequilibrium phases in driven systems with programmable interaction networks.
\end{abstract}

\maketitle
{\it Introduction.---} 
Time-dependent driving offers a versatile toolkit for manipulating quantum many-body systems~\cite{Marin2015,Meinert2016,Eckardt2017,Oka2019} and realizing non-equilibrium phases of matter with no static counterpart, such as discrete time crystals~\cite{Roderich2016,Else2016,yao2017discrete}. Yet, driven many-body systems generally absorb energy and eventually evolve toward a featureless infinite-temperature state~\cite{Prosen1998,Luca2014,Lazarides2014,Huse2014,hou2025floquet}.
A central question is therefore what mechanisms can stabilize periodically driven systems against heating. In many-body systems with local interactions and no strong disorder, the heating rate follows an universal scaling, $\gamma \sim e^{-\mathcal{O}(\omega/J_0)}$, and is exponentially suppressed for driving frequencies $\omega$ much larger than a characteristic local energy scale $J_0$~\cite{Abanin2015,Kuwahara2016,Takashi20161,Abanin2017,Choi2017,Abanin2018,Marin2019,Francisco2019,Philip2019,Anatoli2021,Peng2021}. 
For systems with slowly decaying interactions, heating can be fully suppressed when the underlying mean-field dynamics is non-chaotic~\cite{Pizzi2021Higher,Mu2022Floquetall-to-all,Lerose2025Theory}. 
However, heating in systems with generic long-range interactions remains poorly understood.

Addressing this challenge is particularly timely due to the rapid developments in quantum platforms, where long-range interactions are becoming increasingly programmable. For instance, Rydberg atoms in optical tweezer arrays can be reconfigured to create effective shortcuts between distant sites~\cite{Ebadi2022,Bluvstein2024}, and distant atoms can be coupled in optical cavities via photonic modes~\cite{periwal2021programmable,Marsh2025,grinkemeyer2025error,de2026realization}. These platforms provide an unique opportunity to realize new quantum computing architectures, generate spin squeezing, and hinder decoherence~\cite{kuriyattil2025entangled,gorlach2025supercoherence,solfanelli2026robust}. The complex interaction networks that they can implement are fundamentally different from regular lattices~\cite{Islam2013,Kennes2021,Ebadi2022,Bluvstein2024,Guo2024,Marsh2025}, raising key questions on how heating occurs at a microscopic level and whether it can be controlled by engineering the network topology.

Here, we address these questions and show that the interplay between external driving and interaction network topology can lead to tunable anomalous heating. Our key insight is that heating fundamentally depends on the distribution of the network's coordination numbers, giving rise to a hierarchy of local energy scales and hence resonant driving frequencies, and resulting in anomalous heating phenomena not possible in regular lattices.

\begin{figure*}[htbp]
	\centering
	\includegraphics[width=0.99\textwidth]{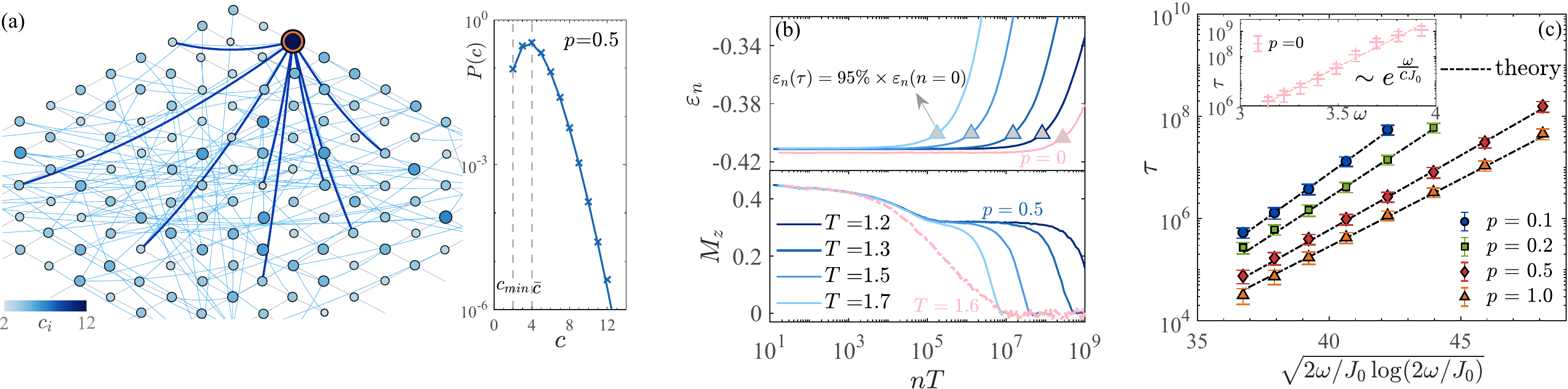}
	\caption{(a) Schematic of SW network and degree distribution. Bonds of a square lattice with nearest-neighbor coordination $2k=4$ are rewired with probability $p\in(0,1]$ to random sites, excluding self-loops and duplicate edges. Node color encodes the coordination number $c$, light-blue arcs indicate rewired shortcuts, and a representative high-degree node together with the shortcuts linked to it are highlighted in dark blue. The coordination number follows the probability distribution $P(c)$. (b) Dynamics of the stroboscopic energy density $\varepsilon_n$ and total magnetization $M_z$ for various driving periods $T$ at fixed $p=0.5$. The SW network heats faster than the regular lattice ($p=0$, pink), yet retains a long-lived prethermal plateau with finite $M_z$. (c) Scaling of the thermalization times $\tau$ for various $p$. The numerics is in excellent agreement with the theory (black-dash-dotted curve), confirming the scaling $\log \tau \sim \sqrt{(\omega/J_0)\ln(\omega/J_0)}$. Inset: Conventional scaling $\log \tau \sim \omega$ for the regular lattice ($p = 0$). For (b) and (c), $L=30$ and $g/J=0.07$. Results are averaged over $200$ network and initial-state realizations sampled with $\delta\theta=0.4\pi$.}
	\label{fig:figtau}
\end{figure*} 

To illustrate this physics we focus on the paradigmatic small-world (SW) networks~\cite{Watts1998},
in which one end of each edge of a regular lattice is ``rewired'' with probability $p$ to a random site, a procedure that naturally mimics the shortcuts enabled by quantum simulators, and in which $p$ provides a knob to control the connectivity and access different heating regimes. We show that even a small fraction of long-range bonds can have dramatic effects on the heating rate, making it scale as $\gamma\sim e^{-\mathcal{O}(\sqrt{\omega/J_0})}$. We match large scale numerics with an analytic theory, where we introduce a notion of \textit{site-dependent heating rate} and show that heating becomes dominated by sites with larger coordination numbers when the driving frequency grows. {This feature is unique to long-range interaction networks and is not restricted to SW networks; it is directly responsible for the anomalous heating.} Finally, we show that this tunable long-lived prethermal regime can stabilize nonequilibrium phases of matter, such as discrete time crystals.

{\it Model.---}
Consider a system of classical spins $\mathbf{S}_i=(S_i^x,S_i^y,S_i^z)$ on a graph $G = (V,E)$, with $V = 1,2,\dots, N$ nodes corresponding to the position of the spins, and $E$ edges corresponding to the interactions between them. The spins are driven by a time-periodic Hamiltonian $H(t)=H(t+T)$,
\begin{equation}
H(t)=
\left\{
\begin{array}{@{}l@{\;}l@{}}
-J\displaystyle\sum\nolimits_{\{i,j\}\in{\rm E}} S_i^zS_j^z,
& nT\le t\le (n+\tfrac12)T,\\[6pt]
g\displaystyle\sum\nolimits_i S_i^x,
& (n+\tfrac12)T<t\le (n+1)T ,
\end{array}
\right.
\label{eq:Ham}
\end{equation}
where $n$ counts the driving cycles and $\omega = 2 \pi/T$ is the drive frequency. 
We consider ferromagnetic coupling $J>0$ and a small transverse field $g\ll J$. 
The spin dynamics reads $\dot{\mathbf{S}}_i = \mathbf{h}_i(t) \times \mathbf{S}_i$, where $\mathbf{h}_i(t) = \partial H(t)/\partial \mathbf{S}_i$ is a local effective field. For the piecewise $H(t)$ in Eq.~\eqref{eq:Ham}, $\mathbf{h}_i(t)$ points along the $z$ or $x$ axis in each half-period, enabling efficient stroboscopic evolution via simple rotation matrices and large-scale simulations over long times~\cite{Marin2019}.
It has been shown that the physics of heating is similar in classical and quantum many-body systems~\cite{Rajak_2018,Das2018,Marin2019,Pizzi2021,Bingtian2021,Guo2025}, and we thus expect our central findings to qualitatively also apply to quantum systems.

To retain significant analytic tractability, we shall primarily focus on SW networks~\cite{Watts1998}. These are obtained from a regular lattice by rewiring each interaction edge with probability $p$, that is, reconnecting one of its ends to a random site; see Fig.~\ref{fig:figtau}(a). Conveniently, SW networks interpolate between a regular lattice ($p=0$) and a random graph ($p=1$). Let us call $k$ the number of edges per site, which is conserved by the rewiring procedure, and $c_i$ the coordination number. The latter is the same for all sites $c_i = c$ in the regular lattice, but becomes site dependent and random after rewiring, with distribution~\footnote{Originally derived for the one-dimensional ring~\cite{Barrat2000}, $P(c)$ is valid irrespective of the dimensionality of the starting regular lattice, provided each site has $k$ outgoing bonds subject to rewiring and incoming rewired links are Poisson distributed in the thermodynamic limit.}
\begin{equation}
P(c) = \sum_{\ell=0}^{\min(c-k,k)} 
\binom{k}{\ell} (1-p)^\ell p^{k-\ell}
\frac{(kp)^{c-k-\ell}}{(c-k-\ell)!} e^{-kp},
\label{eq:degree_distribution}
\end{equation}
where $\ell$ labels the number of original local bonds retained after rewiring; see Fig.~\ref{fig:figtau}(a). 
For concreteness, we construct the network starting from a $L\times L$ two-dimensional regular lattice, hence with $k=2$, but all of our findings can be straightforwardly generalized to networks with other $k$. 

Because the rewiring does not change the total number of edges, the mean coordination number remains $\bar c=2k$, compatibly with Eq.~\eqref{eq:degree_distribution}. Crucially, however, the largest $c_i$ grows with system size. Existing rigorous bounds on the heating rate typically assume a finite number of interaction terms per site~\cite{Takashi2016} -- This assumption breaks here, yielding a rich new physics which we now explore. 

{\it Prethermal dynamics.---} We first investigate the system's dynamics numerically. In spherical coordinates, $\mathbf{S}_i=\big(\sin\theta_i\cos\phi_i, \sin\theta_i\sin\phi_i, \cos\theta_i\big)$, the spins are initialized with $\phi_i$ uniformly random in $(0,2\pi)$ and $\theta_i$ sampled from a Gaussian with zero mean and standard deviation $\delta\theta$. The latter controls the initial energy density with respect to $2H_F^{(0)}=-J\sum_{\{i,j\}\in{\rm E}}S_i^zS_j^z+g\sum_iS_i^x$, which is approximately conserved at early times for high drive frequency $\omega$. A smaller $\delta\theta$ corresponds to lower-energy initial states.

We examine energy absorption by tracking the stroboscopic energy density 
$\varepsilon_n =\langle H_F^{0}(t_n)\rangle/N$ at times $t_n=nT$.  For the regular lattice ($p=0$), the system settles into a long-lived prethermal plateau before notable heating occurs, as shown by the pink curve in the upper panel of Fig.~\ref{fig:figtau}(b). Prethermalization persists for a finite rewiring probability (e.g., $p=0.5$, blue curves), albeit with a shorter lifetime. 
Crucially, the faster heating is not caused by the presence of non-local interaction \textit{per-se}: as we have verified in the Supplemental Material (SM)~\cite{supplement}, heating similar to that of the regular lattices also occurs in random regular graphs, which contain long-range interactions but have a fixed coordination number.

While the rewiring generally accelerates heating, it can stabilize the Ising ferromagnetic order in the prethermal regime. At high frequencies, the stroboscopic dynamics is well approximated by the leading-order Floquet Hamiltonian $H_F^0$, which is a static Ising model on the same network. In 1D, SW connectivity leads to finite-temperature order in $H_F^0$ that is absent in the regular lattice~\cite{Barrat2000}, and in 2D
it raises the critical temperature~\cite{Herrero2002}. The same initial ensemble can then lie in the paramagnetic phase of the regular lattice but in the ordered phase of the SW network after rewiring, with direct consequences on the prethermalization dynamics. This is shown in the lower panel of Fig.~\ref{fig:figtau}(b): the total magnetization $M_z=N^{-1}\sum_i S_i^z$ rapidly decays to zero for the regular lattice (pink), whereas it remains finite throughout the prethermal plateau for the SW network (blue). The dependence of $M_z$ on the rewiring probability and initial energy is detailed in the SM~\cite{supplement}. 

We quantify the heating rate by extracting the prethermal lifetime $\tau$, defined as the time when $\varepsilon_n$ first reaches $0.95$ of its initial value~\footnote{When extracting the lifetime, we use three different tolerance values $0.95-0.01,0.95,0.95+0.01$, which lead to three values of the lifetime. Their average gives the data point in Fig.~\ref{fig:figtau}(c), while the error bar corresponds to their standard deviation.}. For the regular lattice $(p=0)$ and at high frequency $\omega$ we recover the expected exponential scaling $\tau{\sim}\exp( \omega/J_{\mathrm{eff}})$, with $J_{\mathrm{eff}}$ an effective local energy scale, see the inset of Fig.~\ref{fig:figtau}(c). For finite rewiring probability $p$, the prethermal lifetime instead scales as
$\tau\sim\exp(\sqrt{(\omega/J_0)\ln(\omega/J_0)})$, a stretched exponential function with a logarithmic correction, see Fig.~\ref{fig:figtau}(c). Varying $p$ slightly changes the prefactor of the scaling, i.e., the slope of the lines in Fig.~\ref{fig:figtau}(c). Next, we prove this scaling analytically. 

{\it Theory of the heating-rate.---}
The key idea of our analytic theory is to relate the local heating rate with the site-dependent coordination number. Averaging over the distribution in Eq.~\eqref{eq:degree_distribution} leads to the overall heating rate, the black dashed line precisely matching numerics in Fig.~\ref{fig:figtau}(c). Here we outline the main steps of this theory, and details are provided in the SM~\cite{supplement}.

For each site $i$ we consider an energy scale $ J_{\mathrm{eff}}(c_i)$, where $c_i$ denotes the coordination number. For a weak transverse field, this energy is dominated by the Ising interaction and $J_{\mathrm{eff}}(c_i)=J_0c_i$, where $J_0$ estimates the energy scale of each interaction bond. Note, $J_0$ is generally different from the bare Ising interaction strength in Eq.~\eqref{eq:Ham}, and its value can be obtained by numerical fitting the slope of the data shown in the inset of Fig.~\ref{fig:figtau}. For the regular lattice, the coordination number is fixed to $c_i=4$, while for the SW network it is sampled according to the distribution $P(c)$ in Eq.~\eqref{eq:degree_distribution}. 

\begin{figure}[t]
	\centering
	\includegraphics[width=\linewidth]{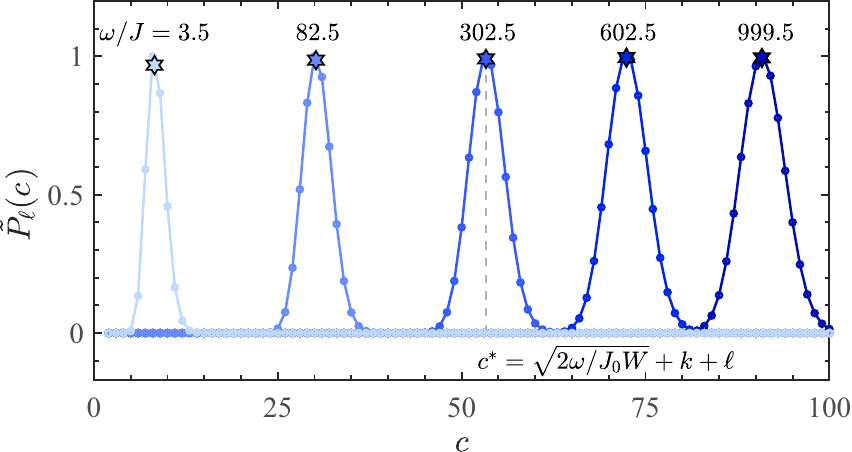}
	\caption{Normalized unimodal contribution $\tilde{P}_\ell(c)$ for fixed $\ell=0$ and $p=0.5$, where each curve is normalized by its own maximum for visibility. In this frequency range, the peak shifts to larger coordination number when the driving frequency $\omega$ increases. Stars mark the saddle-point approximation prediction that correctly captures the position of the peak,  $c^\ast=\sqrt{2\omega/J_0W}+k+\ell$; see main text.}
	\label{fig:2j}
\end{figure}

In spatially homogeneous Floquet systems, given a local energy scale $J_{\text{eff}}$, the heating rate is generally captured by $\gamma=\Gamma_0 e^{-\omega/J_{\mathrm{eff}}}$~\cite{Abanin2015,Kuwahara2016,Takashi20161,Abanin2017,Marin2019,Rigol2019,Anatoli2021}, where $\Gamma_0$ is a prefactor depending on model, driving amplitude, and initial states~\cite{Takashi2022}. In contrast, on a SW network the coordination number varies, and the heating rate becomes explicitly site-dependent, 
\begin{equation}
\label{eq.localrate}
\gamma_i(c_i)=\Gamma_0(c_i)e^{-\omega/J_{\mathrm{eff}}(c_i)}.
\end{equation}
The overall heating rate can thus be obtained averaging as 
$\gamma = \sum_c \gamma_i(c) P(c)$. For low-energy initial states we find $\Gamma_0(c_i)\approx\Gamma_0$, and from Eq.~\eqref{eq:degree_distribution} we get
\begin{equation}
\gamma=\Gamma_0\sum_{\ell=0}^{k=2} f_\ell \sum_{c=k+\ell}^{\infty}\tilde P_\ell(c),
\label{eq:gamma_avg}
\end{equation}
where $\ell$ counts retained local bonds after rewiring, $f_\ell=\binom{k}{\ell}(1-p)^\ell p^{k-\ell}e^{-kp}$, and
\begin{equation}
    \label{eq.P_tilde}
    \tilde P_\ell(c)=\exp\left(-\frac{\omega}{c J_0}\right)\frac{(kp)^{c-k-\ell}}{(c-k-\ell)!}.
\end{equation}
Since $f_\ell$ is independent of $\omega$, the high-frequency scaling is controlled by the $c$-dependent terms $\tilde P_\ell(c)$.

Two factors compete in $\tilde P_\ell(c)$: the exponential increases with $c$, suggesting that a larger coordination number enlarges the local energy scale, opens more heating channels, and speeds up energy absorption. The Poisson-like tail inherited from Eq.~\eqref{eq:degree_distribution} instead decays for $c \to \infty$, and accounts for the fact that large coordination numbers are statistically rare.
Their competition produces a sharply peaked distribution of $\tilde P_\ell(c)$, as shown in Fig.~\ref{fig:2j}. 

The summation over $\tilde P_\ell(c)$ in Eq.~\eqref{eq:gamma_avg} can then be well approximated using a saddle-point approximation, in which, thanks to a Stirling approximation, $\tilde{P}_{\ell}$ is approximated as a continuous Gaussian centered around a $c^*$, that is found numerically. The parameter $\Gamma_0$ is fitted. The resulting prediction is plotted as a dashed line in Fig.~\ref{fig:figtau}(c), in excellent agreement with numerics. The special choice of axes in Fig.~\ref{fig:figtau}(c) makes the data appear in a almost perfect straight line. To understand this scaling, we push the saddle-point approximation one step further by noting that, as shown in Fig.~\ref{fig:2j}, at high frequency $c^\ast \approx \sqrt{2\omega/J_0 W}+k+\ell$, where $W\equiv W(2\omega/J_0 k^2p^2)$ is the Lambert $W$ function defined by $W(x)e^{W(x)}=x$~\footnote{In fact, $\tilde P_\ell(c)$ also broadens when $\omega$ increases, so in principle, the saddle point approximation may fail in the limit $\omega\to\infty$. However, throughout the physically relevant frequency range considered in this work, the saddle-point approximation remains valid.}.
Note that $c^\ast$ increases with $\omega$, indicating that at high frequency heating is dominated by sites with larger coordination numbers. The final overall heating rate is
\begin{align}
\ln\gamma \simeq \ln\Gamma_0 -\sqrt{2\omega W/J_0} .
\label{eq:gamma0}
\end{align}
At high frequencies, the dependence of the Lambert function $W$ on $\omega$ becomes weak, leading to the key finding of our work: in SW networks $\gamma \simeq e^{\mathcal{O}(\sqrt{\omega})}$, in sharp contrast to the standard case $\gamma \simeq e^{\mathcal{O}(\omega)}$ for regular lattices. More precisely, we can further simplify Eq.~\eqref{eq:gamma0}, including prefactors, using $W(x) \simeq \ln x$, which leads to two asymptotic forms depending on the rewiring probability $p$.

\textit{High-frequency regime}: If $\omega/J_0 \gg \max(k^2p^2,\,\frac{1}{k^2p^2})$,
\begin{equation}
\ln\gamma\simeq
\ln\Gamma_0
-\sqrt{2\omega/J_0\ln(2\omega/J_0)}
\left(1 + \alpha \right),
\label{eq:gamma1}
\end{equation}
where $\alpha = - \frac{\ln(kp)}{\ln(2\omega/J_0)}$. This result explains the choice of axes and observed scaling in Fig.~\ref{fig:figtau}(c). The weak dependence on $p$ of the slopes is captured by the subleading correction $\alpha$. 

\textit{Sparse-rewiring regime}: If $p$ is very small but finite, the high-frequency condition $\omega/J_0 \gg 1/(k^2p^2)$ can be hardly satisfied, and our theory predicts  
\begin{equation}
\ln\gamma\simeq
\ln\Gamma_0
-\sqrt{2\omega/J_0\ln(1/k^2p^2)}.
\label{eq:gamma2}
\end{equation}
Simulations in the regime of very small $p$ are challenging because affected by large finite size effects. Yet, the scaling in Eq.~\eqref{eq:gamma2}, which we verified in the SM through careful analysis of the saddle-point approximation, proves a very important conceptual point: \textit{any} finite fractions $p$ of rewired links, no matter how small, turns the standard exponential scaling $\gamma \sim e^{\mathcal{O}(\omega)}$ of regular lattices into an anomalous, stretched exponential $\gamma \sim e^{\mathcal{O}(\sqrt{\omega})}$.

\begin{figure}[t]
	\centering
    \includegraphics[width=0.47\textwidth]{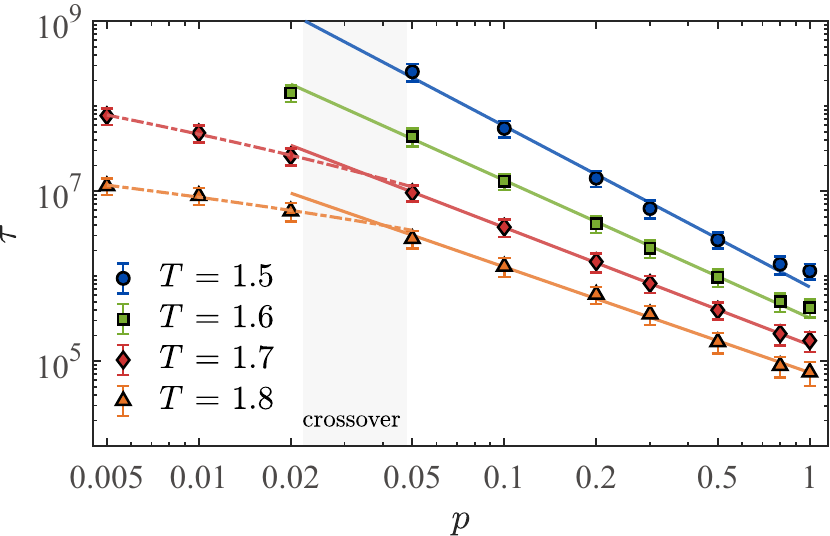}
	\caption{Thermalization time $\tau$ versus rewiring probability $p$ for various driving periods $T$. The data display an algebraic dependence on $p$ in the high-frequency regime (solid fits), before crossing over to the sparse-rewiring regime [dash-dotted fits to Eq.~\eqref{eq:gamma2}]. The shaded region marks the crossover. Simulations use $L=30$--$100$ (larger systems for smaller $p$), and are averaged over $200$ independent network and initial-state realizations at $\delta\theta=0.4\pi$.}
	\label{fig:p}
\end{figure}  

The two regimes can also be appreciated in the $p$ dependence at fixed driving frequency $\omega$, in Fig.~\ref{fig:p}. For $p\gtrsim0.05$, the system is in the high-frequency regime, where the subleading correction $\alpha$ gives an algebraic dependence of $\tau$ on $p$, consistent with the power-law fits (solid lines). As $p$ is reduced, the system enters the sparse-rewiring regime and the data bend down and follow Eq.~\eqref{eq:gamma2} (dash-dotted lines).

{\it Prethermal discrete time crystal.---} The tunable and long-lived prethermal regime can stabilize non-equilibrium phases of matter, e.g., discrete time crystals. To show it, we modify the drive in Eq.~\eqref{eq:Ham} by adding a $\pi/2$ rotation of the spins around the x-axis after each period, $\mathbf{S}_i\mapsto R_x(\pi/2)\mathbf{S}_i$. This can generate period-4 discrete time crystalline order (4-DTC), where the total magnetization of the system follows the evolution
$z\to y\to -z\to -y\to z\to\dots$~\cite{Pizzi2021Higher,Pizzi2021}.

\begin{figure}[t]
	\centering
	\includegraphics[width=0.45\textwidth]{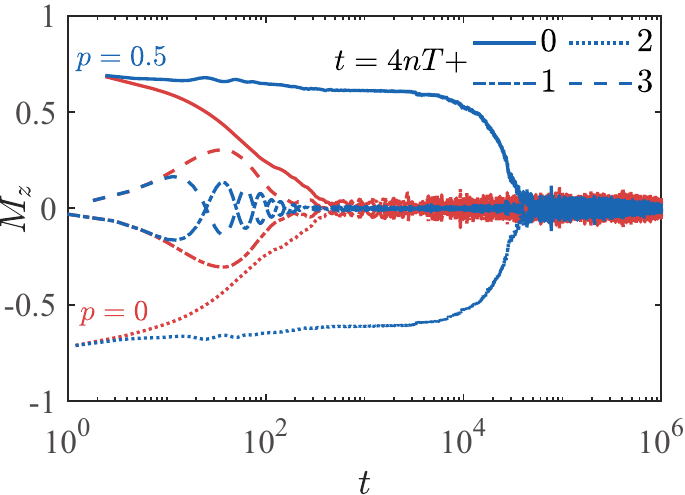}
	\caption{Magnetization dynamics for the regular lattice ($p=0$, red) and the SW network ($p=0.5$, blue). Different line styles distinguish different series of stroboscopic times $t=(4n+m)T$ with $m=0,1,2,3$. The magnetization of the regular lattice rapidly decays to zero, whereas the SW network sustains a robust $4T$ oscillation, indicating a prethermal 4-DTC. Here, $\delta\theta=0.25\pi$ and $L=30$.}
	\label{fig:DTC}
\end{figure} 

Importantly, the emergence of 4-DTC relies on the appearance of the ferromagnetic order of $H_F^{(0)}$ in the prethermal regime, which is favored by the long-range connections. To show this, in Fig.~\ref{fig:DTC} we compare the magnetization in a regular lattice ($p = 0$, red) and in a SW network ($p=0.5$, blue). For the regular lattice, the magnetization $M_z$ rapidly decays to zero, signaling an underlying paramagnetic phase lacking the symmetry-breaking order required for a DTC. For the SW network, instead, long-range shortcuts raise the critical temperature of the prethermal effective Hamiltonian, placing the same initial ensemble in the ferromagnetic phase. 
As a result, $M_z$ exhibits a robust $4T$-periodic oscillation signalling the $4-$DTC, whose lifetime we verified to increase with the driving frequency.

{\it Conclusion and outlook.---}
We have analyzed Floquet heating in SW networks, demonstrating a prethermal regime whose lifetime exhibits an anomalous, stretched-exponential scaling with the driving frequency. This behavior originates from the broad distribution of coordination numbers, which leads to spatially nonuniform local heating rates across the system. We have developed an analytical framework that captures this mechanism and shows how heating at high frequency is dominated by sites with large coordination numbers. Small degrees of long-range connectivity can have drastic effects, e.g., underpinning the celebrated small-world effect in social networks~\cite{milgram1967small}. Remarkably, this applies to heating and prethermalization as well: Eq.~\eqref{eq:gamma2} shows that any small $p>0$ suffices to fundamentally alter heating, leading to an anomalous scaling $\tau \simeq e^{\mathcal{O}(\sqrt{\omega})}$.

Since our key assumption, Eq.~\eqref{eq.localrate}, does not rely on the specific microscopic model, our results should generally
apply to 
other Hamiltonians {and interaction networks,
as supported by the numerics in SM~\cite{supplement}}. It should also apply to quantum many-body systems, although the latter is formidably challenging to simulate.

To the best of our knowledge, stretched-exponential scaling of the prethermal lifetime has not been previously observed in periodically driven systems. Similar behaviors have nevertheless been reported in quasiperiodically driven systems~\cite{Else2020,TM2021,Ye2023}. This raises the intriguing question whether these phenomena can be understood within a unified framework. More generally, our theory suggests that network topology can be used to engineer the $\omega$-dependence of Floquet heating. A systematic classification of heating regimes for different interaction network topologies is therefore an interesting direction for future work.

Our results also provide important insights into stabilizing reconfigurable quantum simulators with programmable interaction networks. Under high-frequency driving, densely connected sites may act as local hot spots that can destabilize the entire system. Exploring possible mechanisms to counteract such local heating events, e.g., using dissipative channels~\cite{Ethan2026,shmalo2026controltransitiontemporallyrandom}, is worth pursuing.

{\it Acknowledgments}.--- 
CYG is supported by the the China Postdoctoral Science Foundation (Grant No. GZC20252199). A.~P acknowledges support from Trinity College Cambridge. HZ is supported by Quantum Science and Technology-National Science and Technology Major Project
(No. 2024ZD0301800) and by the National Natural Science
Foundation of China (Grant No. 12474214), and by High-performance Computing Platform of Peking University.

\makeatletter
\let\oldbibliography\bibliography
\def\bibliography#1{%
    \let\oldaddcontentsline\addcontentsline
    \renewcommand{\addcontentsline}[3]{}
    \oldbibliography{#1}%
    \let\addcontentsline\oldaddcontentsline
}
\makeatother
\bibliography{Floquet_small_world}   
\makeatletter
\let\bibliography\oldbibliography
\makeatother

\clearpage
\onecolumngrid

\begin{center}
\textbf{\large{\textit{Supplementary Material} \\ \smallskip
	for ``Anomalous Floquet Heating from Sparse Long-Range Interactions"}}
\end{center}

\renewcommand{\thesection}{SM\;\arabic{section}}
\setcounter{section}{0}
\renewcommand{\thefigure}{S\arabic{figure}}
\setcounter{figure}{0}
\renewcommand{\theequation}{S.\arabic{equation}}
\setcounter{equation}{0}

\setcounter{secnumdepth}{2}
\tableofcontents

\section{Prethermal magnetization plateau}
In the main text, we briefly noted that the rewiring probability $p$ affects the prethermal magnetization plateau. Here we discuss this effect in more detail.
In the prethermal regime, the plateau is governed by the equilibrium properties of the leading-order Floquet Hamiltonian $H_F^{0}$. Thus, changing $p$ affects the plateau by changing the critical temperature of $H_F^{0}$. The regular two-dimensional lattice has a finite critical temperature $T_c^{\mathrm{reg}}$, while SW rewiring raises it to $T_c(p)>T_c^{\mathrm{reg}}$. For a fixed initial state with effective temperature $T_{\mathrm{init}}$ controlled by $\delta\theta$, this gives three regimes:
\begin{itemize}
    \item If $T_{\text{init}} < T_c^{\text{reg}}$, the regular lattice itself is already ordered, and rewiring further stabilizes the order.
    \item If $T_c^{\text{reg}} < T_{\text{init}} < T_c(p)$, the regular lattice lies in the paramagnetic phase ($M_z\equiv0$), while the SW network sustains a nonzero plateau magnetization.
    \item If $T_{\text{init}} > T_c(p)$, both the regular lattice and the SW network are paramagnetic, and the magnetization vanishes regardless of $p$.
\end{itemize}

\begin{figure}[htbp]
	\centering
	\includegraphics[width=0.98\textwidth]{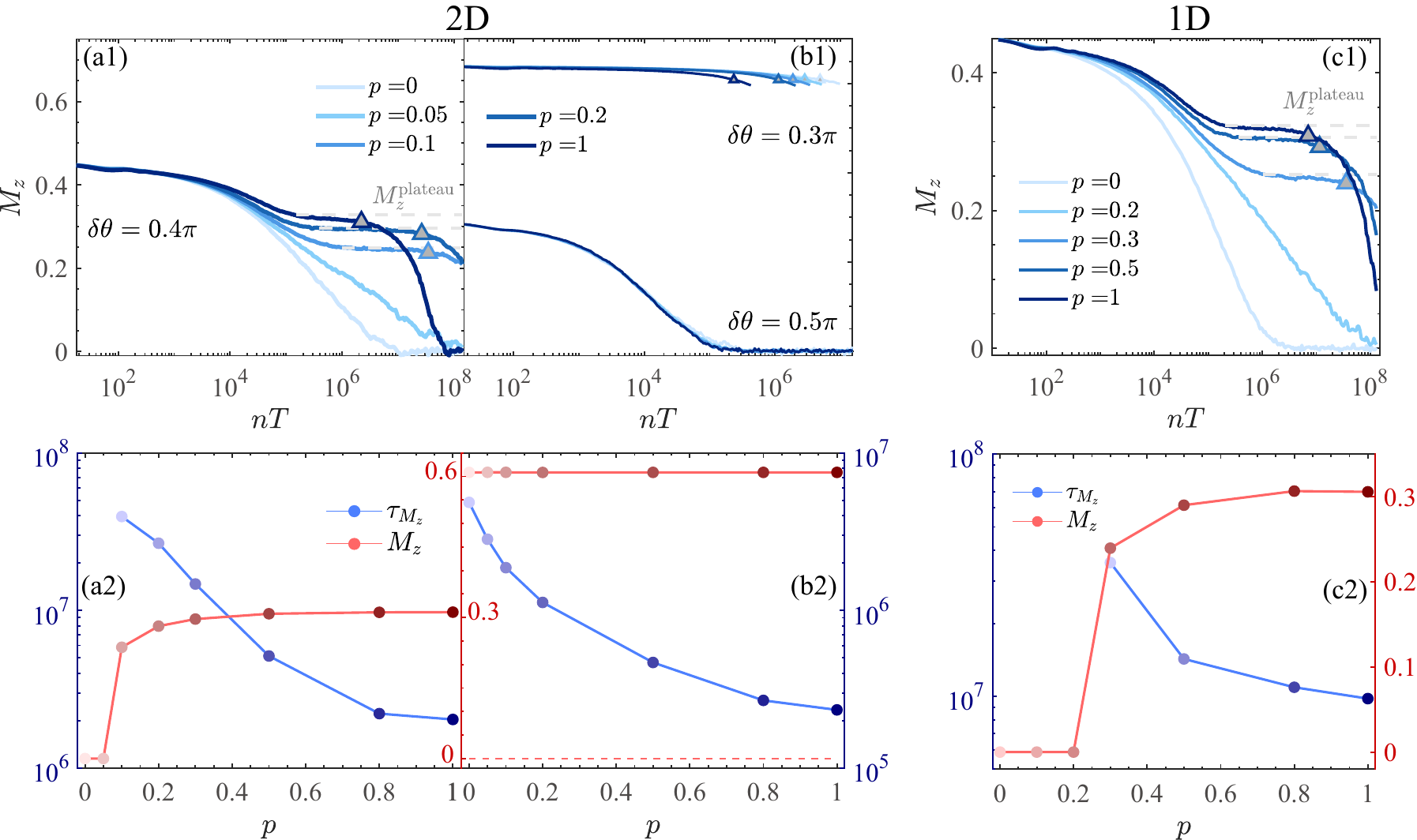}
	\caption{(a1) Dynamics of $M_z$ for various $p$ at $T=1.4J^{-1}$ and $\delta\theta=0.4\pi$, where rewiring produces a finite magnetization plateau absent at $p=0$. (a2) Extracted plateau magnetization $M_z^{\mathrm{plate}}$ (red circles) and magnetic lifetime $\tau_{M_z}$ (blue circles) versus $p$. Together, (a1)--(a2) show that increasing $p$ strengthens the magnetization plateau albeit with a shorter magnetic lifetime. Here $\tau_{M_z}$ is defined as the time when $M_z$  decays to $95\%$ of its plateau value. (b1)--(b2) Corresponding results at $T=1.8J^{-1}$ for ordered ($\delta\theta=0.3\pi$) and disordered ($\delta\theta=0.5\pi$) initial ensembles. For the strongly ordered initial ensemble, increasing $p$ leaves the plateau height nearly unchanged but still shortens its lifetime. (c1)--(c2) Similar behavior are observed in one dimension.} 
	\label{fig:SMMz}
\end{figure}

Our simulations in the main text focus on the second regime $(\delta\theta=0.4\pi)$, where the same initial ensemble gives $M_z=0$ for $p=0$ but a finite plateau for $p>0$, as shown in the Fig.~\ref{fig:SMMz}(a1). At fixed driving period $T=1.4J^{-1}$, increasing $p$ raises the plateau magnetization $M_z^{\rm plateau}$ [Fig.~\ref{fig:SMMz}(a2), right axis], reflecting the enhanced ordering tendency of the SW network. We characterize the lifetime of this plateau by $\tau_{M_z}$, defined as the moment when $M_z$ decays to $95\%$ of its plateau value. As shown in Fig.~\ref{fig:SMMz}(a2) [left axis], $\tau_{M_z}$ decreases with $p$, indicating that the shortcuts that enhance prethermal magnetization also inevitably open additional heating channels that eventually destroy the ordered plateau.

At lower initial energy, $\delta\theta=0.3\pi$, the system is already ordered at $p=0$ [Fig.~\ref{fig:SMMz}(b1)], so increasing $p$ has little effect on $M_z^{\rm plateau}$ [Fig.~\ref{fig:SMMz}(b2), left axis], while still reducing the lifetime $\tau_{M_z}$ [Fig.~\ref{fig:SMMz}(b2), right axis].
At higher initial energy, $\delta\theta=0.5\pi$, the system remains disordered and $M_z$ stays zero for all $p$ considered [Fig.~\ref{fig:SMMz}(b1)].
A similar trend is observed in one dimension [Figs.~\ref{fig:SMMz}(c1)--(c2)].

\section{Derivation of the heating rate in Small World networks}
\label{Derivation}
In this section, we first justify the assumption of a site-independent prefactor $\Gamma_0$ at low initial energies, and then provide the full asymptotic evaluation of the ensemble-averaged heating rate $\gamma$.

\subsection{Site-independent prefactor $\Gamma_0$}
\begin{figure}[t]
	\centering
	\includegraphics[width=0.98\textwidth]{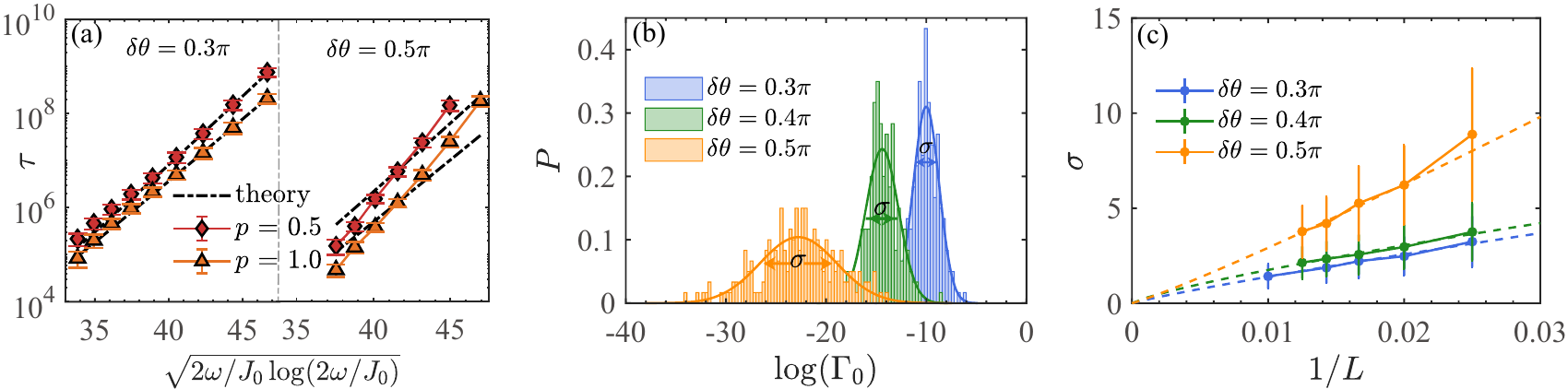}
	\caption{(a) Thermalization time $\tau$ versus $\sqrt{2\omega/J_0 \ln(2\omega/J_0)}$ on a log-linear scale. Numerical data (symbols) agree well with the theoretical prediction (solid blue curve) for the lower-energy initial state $\delta\theta=0.3\pi$ (left), while clear deviations appear for the higher-energy initial state $\delta\theta=0.5\pi$ (right) where $\Gamma_0(c_i)=\Gamma_0$ approximation fails. (b) Probability distribution of the extracted $\ln\Gamma_0$, obtained from $200$ independent realizations of the network and initial state for each $\delta\theta$. The distribution broadens with increasing $\delta\theta$, providing a statistical measure of the increasing sensitivity of $\Gamma_0$ to local environments. (c) Standard deviation $\sigma$ of $\ln\Gamma_0$ versus $L$. Power-law fits $\sigma\sim L^{-\alpha}$ yield $\alpha\approx0.87,0.83,1.04$ for $\delta\theta=0.3\pi,0.4\pi,0.5\pi$, respectively. The values close to $\alpha=1$ are consistent with self-averaging over $N=L^2$ sites, while the larger finite-size value of $\sigma$ at higher $\delta\theta$ reflects stronger local-environment sensitivity. Here, $L=50$ and $p=0.2$.}
	\label{fig:gamma0}
\end{figure}

In the main text, we used the site-dependent heating rate $\gamma_i=\Gamma_0 e^{-\omega/(J_0c_i)}$ under the assumption \begin{equation}
    \label{eq.Gamma}\Gamma_0(c_i)=\Gamma_0.
\end{equation}
 Here we test the range of validity of this approximation by analyzing the heating rate starting from a lower- and higher- energy initial state, with $\delta\theta=0.3\pi$ and $\delta\theta=0.5\pi$, respectively. The main-text simulations are performed at $\delta\theta=0.4\pi$. Fig.~\ref{fig:gamma0}(a) shows that the theoretical prediction agrees well with the numerical data at $\delta\theta=0.3\pi$, while clear deviations appear at $\delta\theta=0.5\pi$.  This suggests that the assumption Eq.~\eqref{eq.Gamma} may not be valid for high-energy initial states.

To validate Eq.~\eqref{eq.Gamma}, one would ideally extract local heating rate $\gamma_i$ at individual sites with a fixed coordination number.  This is difficult in our simulations because the heating time is extracted from global observables, such as the total energy density, which automatically averages over all sites. We therefore test Eq.~\eqref{eq.Gamma} approximation statistically as follows.

For a specific realization of the network and a fixed $\delta\theta$, we extract the global heating time $\tau$ over a range of driving frequencies and fit it to $\ln\tau \simeq \sqrt{2(\omega/J_0)\ln(2\omega/J_0)}-\ln\Gamma_0$, as derived in Eq.~\eqref{eq:gmahigh}. The intercept of this fit gives $\ln\Gamma_0$ for this network realization. Different network realization generally leads to different value of $\ln\Gamma_0$, and the width of its distribution can be used to quantify the fluctuation of the heating rate across different networks. 
A narrow distribution supports the site-independent approximation in Eq.~\eqref{eq.Gamma}, whereas a broad distribution signals stronger dependence of $\Gamma(c_i)$ on the local coordination-number. As shown in Fig.~\ref{fig:gamma0}(b) where the distribution of $\ln\Gamma_0$ is plotted, its width $\sigma$ increases with $\delta\theta$, indicating that $\Gamma_0$ is approximately site independent at relatively low initial energies, but not at higher energies.
We further examine the finite-size scaling of the standard deviation $\sigma$ of $\ln\Gamma_0$ in Fig.~\ref{fig:gamma0}(c). For each $\delta\theta$, $\sigma$ decreases approximately as $\sigma\sim L^{-\alpha}$ with $\alpha\simeq1$. Since $N=L^2$, this is consistent with central limit theorem, $\sigma\sim N^{-1/2}$. The small value of the  slope in  Fig.~\ref{fig:gamma0}(c) for low-energy initial states further validates the assumption in Eq.~\eqref{eq.Gamma}.

\subsection{Derivation of the heating rate}
We now give the detailed derivation of Eqs.~\eqref{eq:gamma1} and ~\eqref{eq:gamma2} in the main text. Using Eq.~\eqref{eq:degree_distribution}, Eq.~\eqref{eq:gamma_avg} can be written as
\begin{equation}
\label{eq:gamma_csum}
\begin{aligned}
\gamma
&=\Gamma_0\sum_{\ell=0}^{k}
\binom{k}{\ell}(1-p)^\ell p^{k-\ell}e^{-kp}
\sum_{c=k+\ell}^{\infty}
\exp\!\left(-\frac{\lambda}{c}\right)
\frac{(kp)^{c-k-\ell}}{(c-k-\ell)!} \\
&=\Gamma_0\sum_{\ell=0}^{k}f_\ell(p)
\sum_{c=k+\ell}^{\infty}\tilde{P}_\ell(c,p) .
\end{aligned}
\end{equation}
where $\lambda=\omega/{J_0}$ and $f_\ell=\binom{k}{\ell}(1-p)^\ell p^{k-\ell}e^{-kp}$, $\ell$ labels the number of original local bonds retained after rewiring. Since $f_\ell(p)$ is independent of $\lambda$ and only sets the overall weight of each $\ell$, the high-frequency scaling is controlled by the remaining summation over $c$. 

As shown numerically in Fig.~\ref{fig:2j} of main text, $\tilde P_\ell(c,p)$ is unimodal and develops a sharp maximum. Thus $c$-dependent sum in Eq.~\eqref{eq:gamma_csum} is dominated by the peak region and can be evaluated by a saddle-point approximation. Note, this approximation requires the peak to be far from the lower boundary of the sum (i.e., $k+\ell$) and narrow compared with the peak position. The first condition gives $\lambda\gg (k+\ell)(k+\ell+1)\ln[1/(kp)]$, which is satisfied in the frequency window considered here; the second condition is verified below after Eq.~\eqref{eq:psi-b}.

For convenience, we shift the summation variable to $j=c-k-\ell$, giving
\begin{equation}
\gamma=\Gamma_0\sum_{\ell=0}^{k}f_\ell(p)S_\ell,\qquad
S_\ell=\sum_{j=0}^{\infty}
\exp\!\left[-\frac{\lambda}{k+\ell+j}\right]\frac{(kp)^j}{j!}
\equiv \sum_{j=0}^{\infty}\tilde P_\ell(j,p).
\label{eq:gamma_sum}
\end{equation}
Treating $j$ as a continuous variable and using Stirling's approximation for
$\psi_\ell(j)\equiv\ln\tilde P_\ell(j,p)$, we obtain
\begin{subequations}
\label{eq:psi}
\begin{align}
\psi_\ell(j) = \ln \tilde{P}_\ell(j,p)
&\simeq -\frac{\lambda}{k+\ell+j} + j\ln(kp) - j\ln j + j - \frac12\ln(2\pi j),\\
\psi_\ell'(j)
&= \frac{\lambda}{(k+\ell+j)^2} + \ln(kp) - \ln j - \frac{1}{2j}, \label{eq:psi-prime}\\
\psi_\ell''(j)
&= -\frac{2\lambda}{(k+\ell+j)^3} - \frac{1}{j} + \frac{1}{2j^2}.
\end{align}
\end{subequations}

Fig.~\ref{fig:2j} in the main text shows that the peak position $c^\ast$, or equivalently $j^\ast=c^\ast-k-\ell$, moves to larger values as frequency increases. We therefore consider the regime $j^\ast\gg k+\ell$, where the finite shift $k+\ell=O(1)$ can be discarded. Then the saddle-point condition $\psi_\ell'(j^\ast)=0$ reduces to $\frac{\lambda}{(j^\ast)^2}\simeq \ln\!\left(\frac{j^\ast}{kp}\right)$, which is independent of $\ell$. Thus all $S_\ell$ in Eq.~\eqref{eq:gamma_sum} share the same leading order contribution to the saddle-point approximation, denoted by $S$ below. It remains to determine the asymptotic scaling of $S$.

Introducing $t=\ln(j^\ast/kp)$, we obtain $j^\ast=kp e^t$ and $t e^{2t}=\lambda/(k^2p^2)$ or $2t e^{2t}=2\lambda/(k^2p^2)$. Hence $2t=W\!\left(\frac{2\lambda}{k^2p^2}\right)$ where $W(x)$ is the Lambert $W$ function defined by $W(x)e^{W(x)}=x$. Therefore,
\begin{equation}
j^\ast=kp\exp\!\left[\frac12 W\!\left(\frac{2\lambda}{k^2p^2}\right)\right]
=\sqrt{\frac{2\lambda}{W\!\left(\frac{2\lambda}{k^2p^2}\right)}}.
\label{eq:j*}
\end{equation}
Evaluating $\psi_\ell$ and $\psi_\ell''$ at $j^*$ yields
\begin{subequations}
\begin{align}
\psi_\ell(j^*) &\simeq -\sqrt{2\lambda W} + \sqrt{\frac{2\lambda}{W}} - \frac12\ln\left(\sqrt{\frac{2\lambda}{W}}\right) - \frac12\ln(2\pi), \label{eq:psi-a}\\
\psi_\ell''(j^*) &\simeq -\frac{W+1}{j^*} = -\frac{W^{3/2}+W^{1/2}}{\sqrt{2\lambda}}. \label{eq:psi-b}
\end{align}
\end{subequations}
Since $\tilde P_\ell(j,p)=e^{\psi_\ell(j)}$, the quadratic expansion of $\psi_\ell(j)$ around the saddle point, $\psi_\ell(j)\simeq \psi_\ell(j^\ast)+\frac{1}{2}\psi_\ell''(j^\ast)(j-j^\ast)^2$, gives a local Gaussian profile with characteristic width $\sigma=1/\sqrt{|\psi_\ell''(j^\ast)|}$. Using Eq.~\eqref{eq:psi-b}, we find $\sigma\ll j^\ast$, which verifies the narrow-peak condition. Therefore the sum can be evaluated by the Gaussian approximation $S\simeq e^{\psi_\ell(j^\ast)}\sqrt{2\pi/|\psi_\ell''(j^\ast)|}$, which reduces to
\begin{equation}
S = (W+1)^{-1/2} \exp\!\left(-\sqrt{2\lambda W} + \sqrt{\frac{2\lambda}{W}}\right).
\end{equation}
Thus Eq.~\eqref{eq:gamma_sum} becomes
\begin{subequations}
\begin{equation}
\gamma \simeq \Gamma_0 e^{-kp} 
\underbrace{\left[ \sum_{\ell=0}^{k} \binom{k}{\ell} (1-p)^\ell p^{k-\ell} \right]}_{=1} S
= \Gamma_0 e^{-kp} S .
\end{equation}
And
\begin{equation}
\label{eq:gmaconst}
\begin{aligned}
\ln \gamma 
&= \log S + \log\Gamma_0 -kp \\
&= -\sqrt{2\lambda} \left(\sqrt{W}+\frac{1}{\sqrt{W}}\right)
-\frac12\ln(W+1) + \ln\Gamma_0-kp \\
&\simeq -\sqrt{2\lambda W} + \ln\Gamma_0
\end{aligned}
\end{equation}
\end{subequations}
We have dropped the small $\lambda$-independent shift $-kp$ and other subleading terms, which are negligible compared with the leading factor $\sqrt{2\lambda W}$ in the parameter range studied here. The suppression of heating is thus controlled by $\sqrt{2\lambda W}$, producing a nonlinear dependence on the driving frequency that contrasts with the regular-lattice result $\ln\gamma\sim-\lambda$. Since $p$ enters through $W(2\lambda/k^2p^2)$, it also introduces an additional control parameter that separates the asymptotics into distinct regimes discussed below.

\begin{figure*}[htbp]
	\centering
	\includegraphics[width=0.8\textwidth]{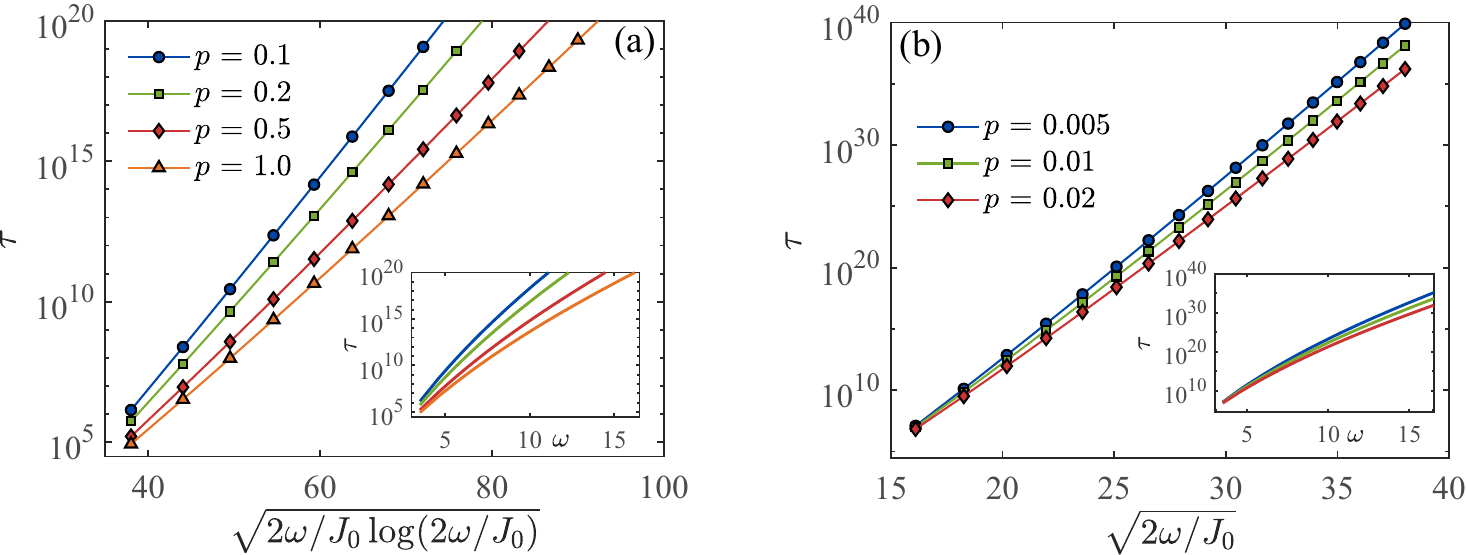}
	\caption{Verification of the two asymptotic scalings in Eqs.~\eqref{eq:gmahigh} and \eqref{eq:gmasp}. The curves are obtained by numerically solving the saddle-point equation for $j^\ast$ and substituting it into Eq.~\eqref{eq:gamma_sum}, with $\tau\sim\gamma^{-1}$. Panels (a) and (b) confirm the predicted scaling behavior in $\sqrt{2\lambda\ln(2\lambda)}$ and $\sqrt{2\lambda}$, respectively. The insets plot the same curves against the bare frequency $\omega$, where both bend downward and are therefore distinct from conventional linear-$\omega$ scaling.}
	\label{fig:SMfig1}
\end{figure*}

To better understand the behavior of the Lambert W function, we use the asymptotic expansion $W(z)\approx\ln z$ for $z=2\lambda/(k^2p^2)\gg1$, and substitute it into Eq.~\eqref{eq:gmaconst}:
\begin{equation}
\ln\gamma \simeq -\sqrt{2\lambda\ln\!\left(\frac{2\lambda}{k^2p^2}\right)
} + \ln\Gamma_0.
\label{eq:gamma_full}
\end{equation}
(i) \textit{\textbf{High-frequency regime}}: $\lambda \gg \max(k^2p^2,\frac{1}{k^2p^2})$. Eq.~\eqref{eq:gamma_full} reduces to
\begin{equation}
\label{eq:gmahigh}
\begin{aligned}
\ln\gamma 
&\simeq -\sqrt{2\lambda\ln(2\lambda)}
+ \sqrt{\frac{2\lambda}{\ln(2\lambda)}}\,\ln(kp)
+ \ln\Gamma_0, \\
&= -\sqrt{2\lambda\ln(2\lambda)}
\left( 1 - \frac{\ln(kp)}{\ln(2\lambda)}\right)
+ \ln\Gamma_0.
\end{aligned}
\end{equation}
The $p$-independent leading term gives the  scaling
$\ln\tau \sim \sqrt{2\lambda\ln(2\lambda)}$, which is a stretched-exponential functional with logarithmic correction.  It captures the numerical scaling law in Fig.~\ref{fig:figtau}(c) of main text. Fig.~\ref{fig:SMfig1}(a) further verifies this asymptotic form: the theoretical curve extrapolated to higher frequencies remains linear when plotted against $\sqrt{2\lambda\ln(2\lambda)}$, while it bends downward when plotted against the bare frequency $\omega$ in the inset. The residual $p$ dependence, visible as the weak variation of the slopes among different $p$ in Fig.~\ref{fig:figtau}(c) of main text, comes from the subleading correction $\sqrt{2\lambda/\ln(2\lambda)}\,\ln(kp)$; relative to the leading term, it is controlled by $\ln(kp)/\ln(2\lambda)$ and therefore vanishes in the high-frequency limit.

At fixed frequency, the subleading correction gives $\tau\propto p^{-\beta(\lambda)}$ with $\beta(\lambda)=\sqrt{2\lambda/\ln(2\lambda)}$. This power-law is in accordance with the numerical results shown in Fig.~\ref{fig:p}. For smaller $p$, the condition $\lambda \gg 1/(k^2p^2)$ breaks down, leading to the sparse-rewiring regime discussed as follows.

(ii) \textit{\textbf{Sparse-rewiring regime:}} $(k+\ell)(k+\ell+1)\ln\frac{1}{kp} \ll \lambda \ll \frac{1}{k^2p^2}$, i.e., $kp \ll 1$. Eq.~\eqref{eq:gamma_full} then simplifies to
\begin{equation}
\label{eq:gmasp}
\begin{aligned}
\ln\gamma
&\simeq -\sqrt{2\lambda\ln\!\Bigl(\frac{1}{k^2p^2}\Bigr)}
+\frac{1}{2}\sqrt{\frac{2\lambda}{\ln(1/(k^2p^2))}}\ln(2\lambda)
+\ln\Gamma_0 \\
&\simeq -\sqrt{2\lambda\ln\!\Bigl(\frac{1}{k^2p^2}\Bigr)}+\ln\Gamma_0 .
\end{aligned}
\end{equation}
It exhibits a stretched-exponential suppression of heating, $\gamma\sim e^{-\mathcal{O}(\sqrt{2\lambda})}$, or equivalently $\tau\sim e^{\mathcal{O}(\sqrt{2\lambda})}$, which remains distinct from the linear-$\omega$ behavior at $p=0$, as evidenced in Fig.~\ref{fig:SMfig1}(b). 

In the sparse-rewiring regime, direct many-body Floquet simulations become particularly demanding because extremely sparse networks require much larger system sizes to obtain converged ensemble-averaged heating times. This numerical difficulty should be distinguished from the saddle-point analysis itself, which remains well controlled in this regime and yields Eq.~\eqref{eq:gmasp}, as illustrated in Fig.~\ref{fig:SMfig1}(b). At the smallest accessible rewiring probability, ($p=0.005$), the available direct-dynamics data are limited to two frequency points, which are sufficient to fix the prefactor ($\Gamma_0$). With this prefactor fixed, the data are compatible with the sparse-rewiring form in Eq.~\eqref{eq:gmasp}, supporting a singular distinction between any finite $p$ and the regular-lattice limit $p=0$.

\section{Heisenberg-type driving protocols}
Here we show that the stretched exponential scaling reported in the main text can also appear in other Hamiltonian systems.
Concretely, we consider the driving protocol
\begin{equation}
	H(t)=
	\begin{cases}
			-J\sum_{\{i, j\}\in \rm E} S_i^x S_j^x;  &  nT <t\leq (n+\frac 13) T\\
			-J\sum_{\{i, j\}\in \rm E} S_i^y S_j^y;  & ( n+\frac 13) T <t\leq( n+\frac 23) T\\
		    -J\sum_{\{i, j\}\in \rm E} S_i^z S_j^z;  &  (n+\frac 23) T <t\leq (n+1) T
	\end{cases} \label{eq:Ham2}
\end{equation}
The sum over $\{i, j\}$ runs over all edges of the SW network.

\begin{figure}[htbp]
	\centering
	\includegraphics[width=0.90\textwidth]{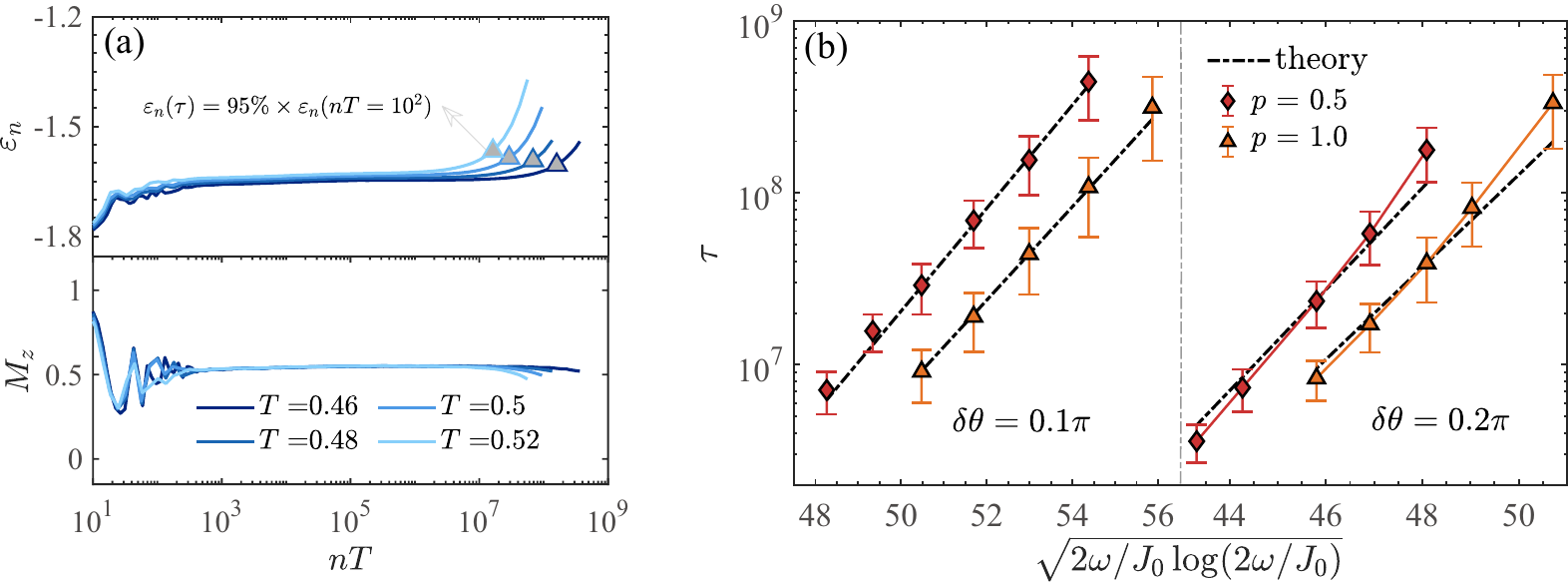}
	\caption{(a) Dynamics of the stroboscopic energy density $\varepsilon_n$ and total magnetization $M_z$ for various driving periods $T$ at fixed $p=0.5$. Darker shades of blue correspond to higher frequencies ($\omega=2\pi/T$). The energy first relaxes to a prethermal plateau and then heats at later times. (b) Thermalization time $\tau$ versus $\sqrt{2\lambda \ln(2\lambda)}$ on a log-linear scale. Here $\tau$ is defined by $\varepsilon_n$ reaching $0.95\pm0.01$ of its plateau value. At low initial energy ($\delta\theta=0.1\pi$, left), the data are well captured by the theoretical prediction, while deviations appear at $\delta\theta=0.2\pi$ (right), consistent with the Ising drive.}
	\label{fig:Heisenberg}
\end{figure} 

Fig.~\ref{fig:Heisenberg}(a) shows the energy density $\varepsilon_n$ and magnetization $M_z$ on the SW network ($p=0.5$) under the Floquet Heisenberg driven protocol [Eq.~\eqref{eq:Ham2}], with the initial states constructed as in the main text and controlled by $\delta\theta$. After a short initial relaxation, the energy settles into a long-lived prethermal plateau before the eventual heating, and the plateau lifetime increases with $\omega$. We define the thermalization time $\tau$ as the time when $\varepsilon_n$ first reaches $0.95\pm0.01$ of this plateau value.
Fig.~\ref{fig:Heisenberg}(b) shows the extracted $\tau$ versus $\sqrt{2\lambda\ln(2\lambda)}$. At low initial energy ($\delta\theta=0.1\pi$, left panel), the data are well captured by the theoretical prediction, while at higher energy ($\delta\theta=0.2\pi$, right panel) deviations appear. This trend is consistent with the behavior found for the Ising drive.
Our theory works nicely for low-energy initial states and it would be worthwhile to further develop a theory to better capture the heating rate for high-energy initial states.

\section{Other networks}
In this section we examine several different interaction networks. Section A considers a random regular graph, which allows us to study Floquet heating in systems with random connections but a fixed coordination number. Section B 
considers an exponential-distribution network where our theory equally applies.

\subsection{Random regular graph}
We consider a random regular graph (RRG), where each site has exactly four neighbors but the connections are randomized. This can be generated by starting from the regular lattice, then we we uniformly select pairs of bonds and swap their endpoints; the swap probability $p_{\rm swap}$ controls the fraction of randomized edges. This procedure preserves the coordination number $c_i=2k$ for all sites while introducing long-range connections.

\begin{figure}[htbp]
	\centering
	\includegraphics[width=0.80\textwidth]{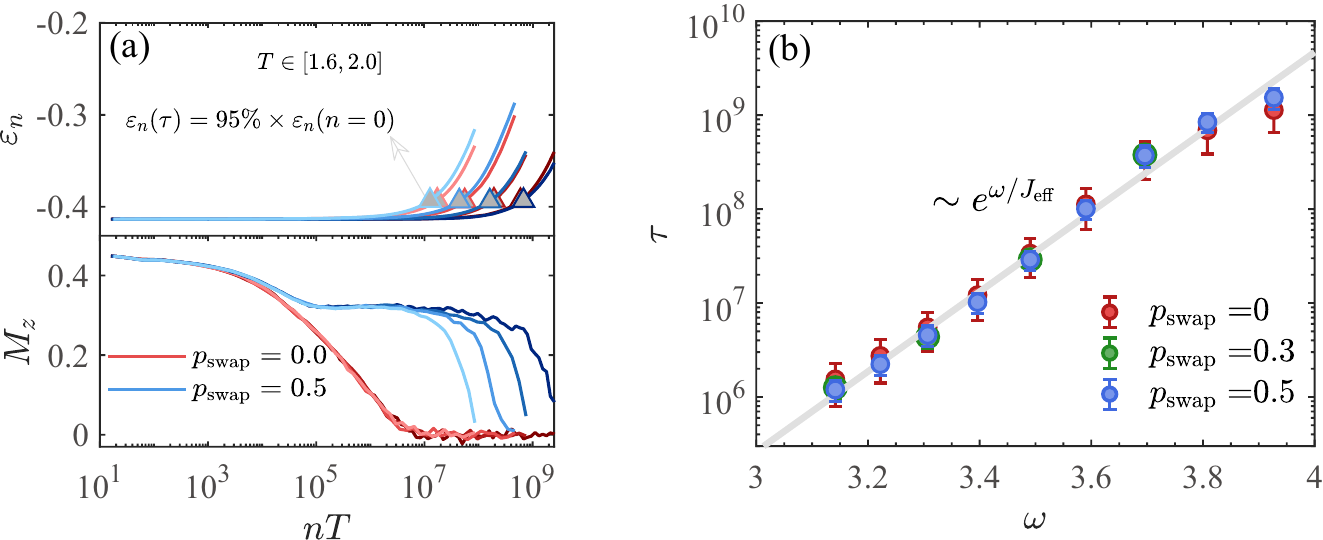}
	\caption{(a) Energy density $\varepsilon_n$ (upper) and total magnetization $M_z$ (lower) on the regular lattice ($p=0$) and on the RRG at $p_{\text{swap}}=0.5$. The energy curves are nearly identical within numerical uncertainty, while $M_z$ vanishes for $p=0$ but remains ordered on the RRG. (b) Heating time $\tau$ extracted from the energy curves for $p_{\text{swap}}=0,0.3,0.5$. All follow the conventional exponential scaling in $\omega$. Parameters are identical to those in Fig.~\ref{fig:figtau} of the main text: $J=1$, $g=0.07$, $\delta\theta=0.4\pi$, $L=30$.}
	\label{fig:RRg}
\end{figure} 

Fig.~\ref{fig:RRg}(a) compares the energy density $\varepsilon_n$ (upper panel) on the regular lattice and on the RRG at $p_{\text{swap}}=0.5$. The two energy curves are nearly indistinguishable and small deviations appear only at late times. The total magnetization $M_z$ (lower panel), however, behaves differently: it rapidly vanishes on the regular lattice but remains ordered on the RRG. Thus, while the randomized rewiring in the RRG do not alter the energy absorption dynamics, they suffice to stabilize ferromagnetic order. We quantify this heating dynamics by extracting $\tau$ from the energy curves and plotting it against $\omega$ in Fig.~\ref{fig:RRg}(b) for several values of $p_{\text{swap}}$. Since bond swapping preserves a fixed coordination number, the local energy scale remains bounded and essentially uniform in space. Thus the RRG falls within the conventional high-frequency regime proved in Ref.~\cite{Takashi2016}, where $\tau\sim e^{\mathcal{O}(\omega)}$. Fig.~\ref{fig:RRg}(b) numerically confirms this behavior: $\tau$ is nearly independent of $p_{\text{swap}}$, and $\ln\tau$ is linear in $\omega$.

These results provide crucial insights of the anomalous heating as seen in SW networks. Randomized long-range connections alone are not sufficient for the anomalous behavior, since the RRG heats similarly to the regular lattice. The essential ingredient is the site-to-site variation of the coordination number: unlike the RRG with fixed $c_i=4$, the SW network has a broad distribution $P(c)$, enabling the frequency-dependent selection of dominant heating channels.

\subsection{Exponential-distribution networks}
In this section, we demonstrate that our heating mechanism applies beyond the small-world ensemble. To this end, we consider a class of sparse networks with exponentially distributed coordination numbers.

The network is generated by a simple modification of the small-world rewiring protocol. In the standard construction, each rewired link chooses its new endpoint uniformly among all sites. Here, each site $i$ is instead assigned with an independent positive weight $u_i$ drawn from an exponential distribution, and a rewired link chooses site $i$ with probability proportional to $u_i$. The resulting coordination-number distribution takes a form analogous to that of the standard small-world ensemble, with the Poisson factor replaced by an exponential one:
\begin{equation}
P_{\rm exp}(c)=\sum_{\ell=0}^{\min(c-k,k)}
\binom{k}{\ell}(1-p)^\ell p^{k-\ell}
\frac{1}{1+kp} e^{-b_p(c-k-\ell)},
\qquad c\ge k ,
\end{equation}
where $b_p=\ln[(1+kp)/(kp)]$.
Following the same averaging procedure as in Sec.~\ref{Derivation}, Eq.~\eqref{eq:gamma_csum} becomes, for the present ensemble,
\begin{equation}
\begin{aligned}
\gamma
&=\frac{\Gamma_0}{1+kp}
\sum_{\ell=0}^{k}
\binom{k}{\ell}(1-p)^\ell p^{k-\ell}
\sum_{c=k+\ell}^{\infty}
\exp\left[-\frac{\lambda}{c}-b_p(c-k-\ell)\right] \\
&=\frac{\Gamma_0}{1+kp}
\sum_{\ell=0}^{k}
\binom{k}{\ell}(1-p)^\ell p^{k-\ell}S_\ell ,
\end{aligned}
\label{eq:gamma_exp_avg}
\end{equation}
where
\begin{equation}
S_\ell=\sum_{j=0}^{\infty}
\exp\left[-\frac{\lambda}{k+\ell+j}-b_pj\right].
\label{eq:Sell_exp}
\end{equation}
Similarly, using a saddle-point approximation, the peak is located at $j^\ast=\sqrt{\frac{\lambda}{b_p}}-k-\ell$. Substituting the saddle-point estimate into Eq.~\eqref{eq:gamma_exp_avg} gives
\begin{equation}
\ln\gamma\simeq \ln\Gamma_0 - 2\sqrt{\lambda b_p} 
=\ln\Gamma_0 - 2\sqrt{\ln\frac{1+kp}{kp}}\sqrt{\frac{\omega}{J_0} } .
\label{eq:gamma_exp}
\end{equation}
It means that the exponential-distribution network also exhibits stretched-exponential suppression of heating. We now compare this asymptotic prediction with direct numerical simulations.

\begin{figure}[htbp]
	\centering
	\includegraphics[width=0.95\textwidth]{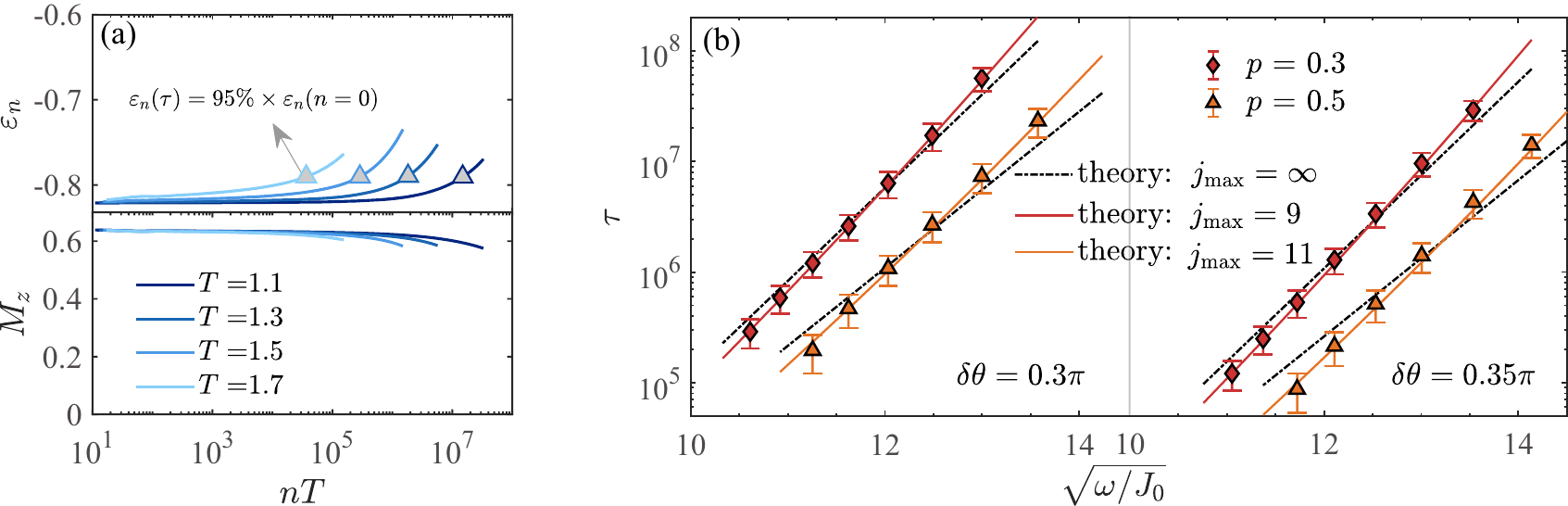}
	\caption{(a) Stroboscopic dynamics of the energy density $\varepsilon_n$ and total magnetization $M_z$ for several driving periods $T$ at fixed $p=0.5$ and $\delta \theta=0.3\pi$. Darker shades of blue correspond to higher driving frequencies $\omega=2\pi/T$. The energy first enters a long-lived prethermal plateau and heats only at later times, and the magnetization remains finite over the prethermal window. (b) Thermalization time $\tau$ versus $\sqrt{\omega/J_0}$ on a log-linear scale for two low-energy initial states, $\delta\theta=0.3\pi$ and $0.35\pi$. Symbols denote numerical data for $p=0.3$ and $p=0.5$. The dashed curves show the asymptotic saddle-point theory, while the solid curves include the finite-network cutoffs $j_{\max}=9$ for $p=0.3$ and $j_{\max}=11$ for $p=0.5$. Here, $L=100$ and $g/J=0.07$. Results are averaged over $200$ network and initial-state realizations sampled with $\delta\theta$.}
	\label{fig:SM_exp}
\end{figure} 

Fig.~\ref{fig:SM_exp}(a) shows representative stroboscopic dynamics. As in the standard small-world case, the energy $\varepsilon_n$ first enters a long-lived prethermal plateau; increasing the driving frequency extends the plateau lifetime. The magnetization $M_z$ remains finite over the prethermal window and decays only during the subsequent heating process.

The extracted heating times are shown in Fig.~\ref{fig:SM_exp}(b) for two different low-energy initial states and they follow the stretched-exponential functional form of Eq.~\eqref{eq:gamma_exp}. However, clear deviations from the theoretical prediction (black dashed line), obtained using the aforementioned saddle-point approximation, appear. We attribute this discrepancy to the finite-size effect as explained in the following. 

We first note that the saddle point $j^\ast$ selects the dominant contribution to Eq.~\eqref{eq:Sell_exp}
by assuming a thermodynamically large system where $j$ can be infinitely large.
However, 
in a finite network, this assumption may not work and hence the  saddle-point approximation can fail. 
To account for this finite-size effect, we therefore replace the infinite sum in Eq.~\eqref{eq:Sell_exp} by
\begin{equation}
S_\ell^{(j_{\max})}=
\sum_{j=0}^{j_{\max}}
\exp\left[-\frac{\lambda}{k+\ell+j}-b_pj\right],
\label{eq:Sell_exp_cut}
\end{equation}
and use $S_\ell^{(j_{\max})}$ in Eq.~\eqref{eq:gamma_exp_avg}. The cutoff $j_{\max}$ is estimated from the finite-size network coordination number  statistics. Since $j=c-k-\ell$, the largest coordination number in the generated samples set the relevant cutoff in $j$. For the $L=100$ networks in Fig.~\ref{fig:SM_exp}(b), we use typical values $j_{\max}\simeq 9$ for $p=0.3$ and $j_{\max}\simeq 11$ for $p=0.5$. With these finite-size cutoffs, the truncated theory agrees well with the numerical data, as shown by the solid curves in Fig.~\ref{fig:SM_exp}(b).

For comparison,
this finite-size correction is much less visible in the small-world ensemble, where the degree distribution decays more rapidly as the coordination number grows. In the exponential-distribution network, the slower decay of the coordination number tail makes the cutoff effect more apparent. Nevertheless, after accounting for this finite-size cutoff, the numerics are well described by the truncated theory and still exhibit stretched-exponential heating. This confirms that the heating mechanism driven by coordination-number fluctuations applies beyond the small-world networks.

\end{document}